# Distinct domain-wall motion between creep and flow regimes near the angular momentum compensation temperature of ferrimagnet

Yuushou Hirata,[1,†] Duck-Ho Kim,[1,†,★] Takaya Okuno,[1] Woo Seung Ham,[1] Sanghoon Kim,[1] Takahiro Moriyama,[1] Arata Tsukamoto,[2] Kab-Jin Kim,[3,★] and Teruo Ono[1,4,★]

[1]Institute for Chemical Research, Kyoto University, Uji, Kyoto 611-0011, Japan.

[2]College of Science and Technology, Nihon University, Funabashi, Chiba 274-8501, Japan.

[3]Department of Physics, Korea Advanced Institute of Science and Technology, Daejeon 34141, Republic of Korea.

[4]Center for Spintronics Research Network (CSRN), Graduate School of Engineering Science, Osaka University, Machikaneyama 1-3, Toyonaka, Osaka 560-8531, Japan.

†These authors contributed equally to this work.

★Correspondence to: kim.duckho.23z@st.kyoto-u.ac.jp, kabjin@kaist.ac.kr, ono@scl.kyoto-u.ac.jp

We investigate a magnetic domain-wall (DW) motion in two dynamic regimes, creep and flow regimes, near the angular momentum compensation temperature ($T_A$) of ferrimagnet. In the flow regime, the DW speed shows sharp increase at $T_A$ due to the emergence of antiferromagnetic DW dynamics. In the creep regime, however, the DW speed exhibits a monotonic increase with increasing the temperature. This result suggests that, in the creep regime, the thermal activation process governs the DW dynamics even near $T_A$. Our result unambiguously shows the distinct behavior of ferrimagnetic DW motion depending on the dynamic regime, which is important for emerging ferrimagnet-based spintronic applications.

The rare earth (RE)–transition metal (TM) compounds, in which RE and TM moments are coupled antiferromagnetically, is receiving of great attention because they have two unique compensation temperatures [1–10]. One is the magnetization compensation temperature $T_M$ [3, 4, 11], at which the total magnetic moment goes zero. The other is the angular momentum compensation temperature $T_A$, at which the net angular momentum vanishes [2–4, 12]. The $T_A$ is receiving of particular interest because of the possibility to have a fast antiferromagnetic spin dynamics. Recent experiment has indeed shown the fast DW dynamics at $T_A$ [12], which highlights an important role of $T_A$ in the dynamics of ferrimagnetic DW. To extend the knowledge of ferrimagnetic DW dynamics near $T_A$, here we investigate a temperature dependence of the DW motion in two different dynamic regimes: the creep and flow regimes.

For this study, ferrimagnetic amorphous GdFeCo films with perpendicular magnetic anisotropy (PMA) were prepared. 5-nm SiN (top)/30-nm $Gd_{23}Fe_{67.4}Co_{9.6}$/5-nm Cu/5-nm SiN (bottom) films were deposited on Si wafers using DC magnetron sputtering. The GdFeCo film is then patterned into microstrips having 3-$\mu$m width, 100-$\mu$m length, and 3-$\mu$m wide Hall bar structures using photolithography and Ar ion milling [13]. The distance between Hall bars were set to 30 $\mu$m. 5-nm Ti/100-nm Au electrodes are stacked onto the ends of the wire and Hall bars for current injection and Hall measurement [see Fig. 1(a)].

We first try to determine the magnetization compensation temperature of GdFeCo microstrip. To this end, the anomalous Hall effect (AHE) resistance $R_H$ ($\equiv V_H/I$) is measured with respect to the magnetic field, $B_z$, for various temperatures (160 K < $T$ < 170 K with 1-K interval). The inset of Fig. 1(b) shows the typical result of $R_H$ when we sweep the magnetic field $B_z$. A square hysteresis loop is observed for all temperatures examined, indicating the perpendicular magnetic anisotropy of the patterned GdFeCo microstrip. The magnitude of the AHE resistances, which is defined as $\Delta R_H \equiv R_H(B_z > B_c) - R_H(B_z < -B_c)$ with coercive field $B_c$, are summarized in Fig. 1(b). The $\Delta R_H$ shows a sign change at $T \sim 165.5$ K, indicating that the $T_M$ is approximately 165.5 K [11, 12].

The DW dynamics above the $T_M$ next can be investigated, where $T_A$ is expected to appear [12]. To measure the DW speed, we adopt the real-time DW measurement technique as described in the following [13, 14]. The GdFeCo microstrip is first saturated by a sufficiently large magnetic field ($B$ = -150 mT) to the downward direction and then, a magnetic field ($B_z$), which is smaller than the coercive field ($B_C$) but is larger than propagation field ($B_P$) is applied to the upward direction. Note that the $B_z$ does not create DWs nor reverse the magnetization because the $B_z$ is smaller than $B_C$. Subsequently, a current pulse (12–16 V and 5–30 ns) is injected into the left vertical electrode as shown in Fig.1(a), which creates a DW near the electrode. Once the DW is created, the DW is immediately moved by $B_z$, because the $B_z$ is larger than the $B_P$. When the DW passes through the Hall cross, the Hall voltage drop is recorded in the oscilloscope, from which we obtain the arrival time $t$. The DW speed $v$ is then calculated by the travel length $l$ and the arrival time $t$. The DW speed was determined from 10 times repeated measurements for each $B_z$. The temperature ranging from 200 K to 300 K is examined using the low temperature probe station.

To define the dynamic regime of DW, we investigate the magnetic field dependence of DW speed $v$. Figure 2(b) shows the $v$ - $B_z$ relation obtained at $T = 260$ K. Threshold magnetic field ($B_z^{th} \sim 30$ mT) is clearly observed, suggesting that the thermally activated creep DW motion can appear near $B_z^{th}$ [15–22]. For larger magnetic field ($B_z > 40$ mT), on the other hand, the DW velocity shows linear increase by satisfying $v = \mu B_z$. Here $\mu$ is the DW mobility. This suggests that DW motion belongs to the flow regime in the higher magnetic field [13, 20, 23–25]. Therefore, the magnetic field dependence allows us to investigate the DW dynamics in two different dynamic regimes. We confirmed that the DW speed shows a similar field dependence at all temperatures examined.

The flow regime is first investigated. Figure 2(b) shows $v$ with respect to $T$ for $B_z = 50$ mT. The $v$ exhibits a maximum at $T \sim 240$ K as indicated by the blue arrow. This result is in line with the recent observation that the DW speed becomes maximized at the angular moment compensation temperature $T_A$ due to the pure antiferromagnetic spin dynamics at $T_A$ [12]. Therefore, we can conclude that the $T_A \sim 240$ K in our GdFeCo microstrip.

An important outstanding question is whether the DW speed exhibits sharp increase at $T_A$ even in the creep regime. To check this, we perform the experiment near $B_z^{th}$ for $T > T_A$. Figure 3(a) shows the $\log t$ with respect to temperature for several magnetic fields. Blue and red symbols correspond to the data in creep ($B_z < 40$ mT) and flow regime ($B_z > 40$ mT). Here, the reason why we plot the $\log t$ instead of $\log v$ is that it is hard to define the creep velocity due to stochasticity (that is, measured $t$ may not be the arrival time but be the depinning time). The result shows that the temperature dependence of $t$ is clearly different depending on the dynamic regime. To clearly see the difference, we define a slope of $\log(t)$-$T$ as $\beta$ ($\equiv \log(t)/T$). Figure 3(b) summarizes $\beta$ with respect to $B_z$. It is clear that $\beta$ is positive in the flow regime ($B_z > 40$ mT). This means that, in the flow regime, the DW velocity decreases with increasing the temperature for $T > T_A$, as observed in Fig. 2(b). Contrary to this, $\beta$ has a negative value in the creep regime ($B_z < 40$ mT). That is, in the creep regime, the higher the temperature, the shorter (faster) the DW depinning time (speed). This result is consistent with the thermal activation process, in which the depinning time decreases with increasing temperature due to the assistance of the thermal energy. This means that the unique antiferromagnetic DW dynamics observed at $T_A$ is not relevant in the DW creep regime. Instead, the thermal activation over energy barriers dominates the DW motion in the creep regime. Our results therefore imply that the identification of the dynamic regime is important for ferrimagnet-based spintronic applications [26–29].

In conclusion, we have investigated the motion of ferrimagnetic DW near the angular momentum compensation temperature in two dynamic regimes, creep and flow regimes. We found a distinct temperature dependence of the DW speed between two dynamic regimes. The DW speed shows a peak at $T_\text{A}$ in the flow regime, whereas it increases monotonically with increasing temperature in the creep regime. These observations imply that the DW dynamics is governed by the total angular momentum in flow regime, whereas it is dominated by the thermal activation process in creep regime. Our findings therefore suggest that the identification of DW dynamic regime is important for emerging ferrimagnet-based spintronic applications.

**Figure Captions**

**Figure 1.** (a) Optical image of the device structure with schematic illustration of the measurement set-up for real-time domain wall (DW) motion. (b) The magnitude of anomalous Hall resistance ($\Delta R_\text{H}$) as a function of temperature ($T$). The red arrow indicates the magnetization compensation temperature ($T_\text{M}$). The inset shows the $R_\text{H}$ with respect to the magnetic field ($B_\text{z}$) at $T = 170$ K.

**Figure 2.** (a) DW speed $v$ with respect to $B_\text{z}$ at $T = 260$ K. The blue box indicates the creep regime. The red dotted line represents the best linear fit based on $v = \mu B_\text{z}$. (b) DW speed $v$ as a function of $T$ for $B_\text{z} = 50$ mT. The red arrow represents $T_\text{M}$ and the blue arrow indicates $T_\text{A}$.

**Figure 3.** (a) The measured $\log(t)$ as a function of $T$ for several $B_\text{z}$. (b) The slope in Fig. 3(a) which is defined as $\beta(\equiv \log(t)/T)$ with respect to $B_\text{z}$. The dotted lines guide the eye.

**Acknowledgements**

This work was partly supported by JSPS KAKENHI Grant Numbers 15H05702, 26870300, 26870304, 26103002, 25220604, 2604316 Collaborative Research Program of the Institute for Chemical Research, Kyoto University, and R & D project for ICT Key Technology of MEXT from the Japan Society for the Promotion of Science (JSPS). D.-H.K. was supported from Overseas Researcher under Postdoctoral Fellowship of JSPS (Grant Number P16314). KJK was supported by the National Research Foundation of Korea (NRF) grant funded by the Korea government (MSIP) (No. 2017R1C1B2009686) and by the DGIST R&D Program of the Ministry of Science, ICT and Future Planning (17-BT-02).

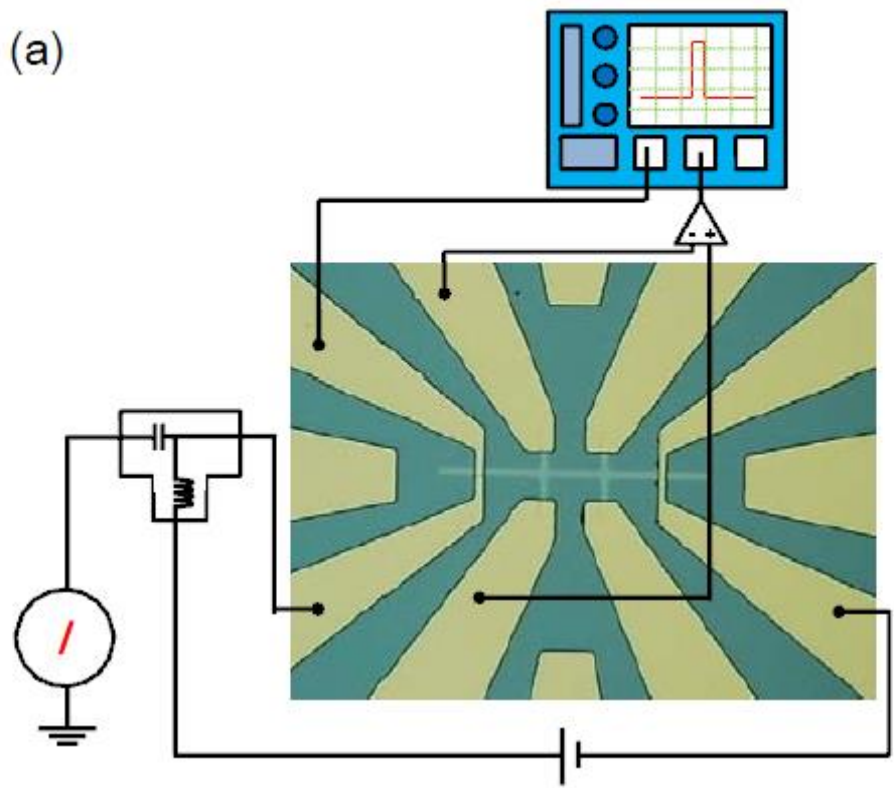

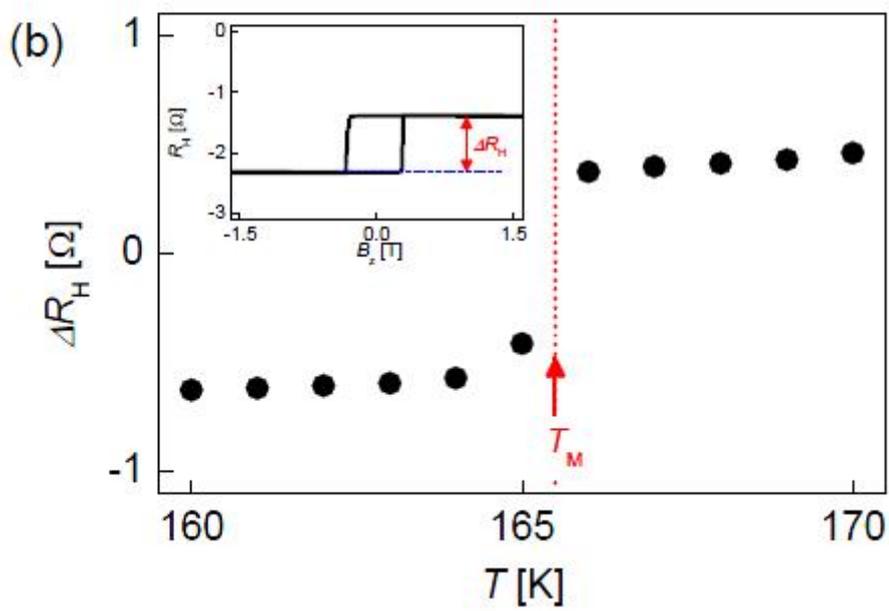

Fig. 1

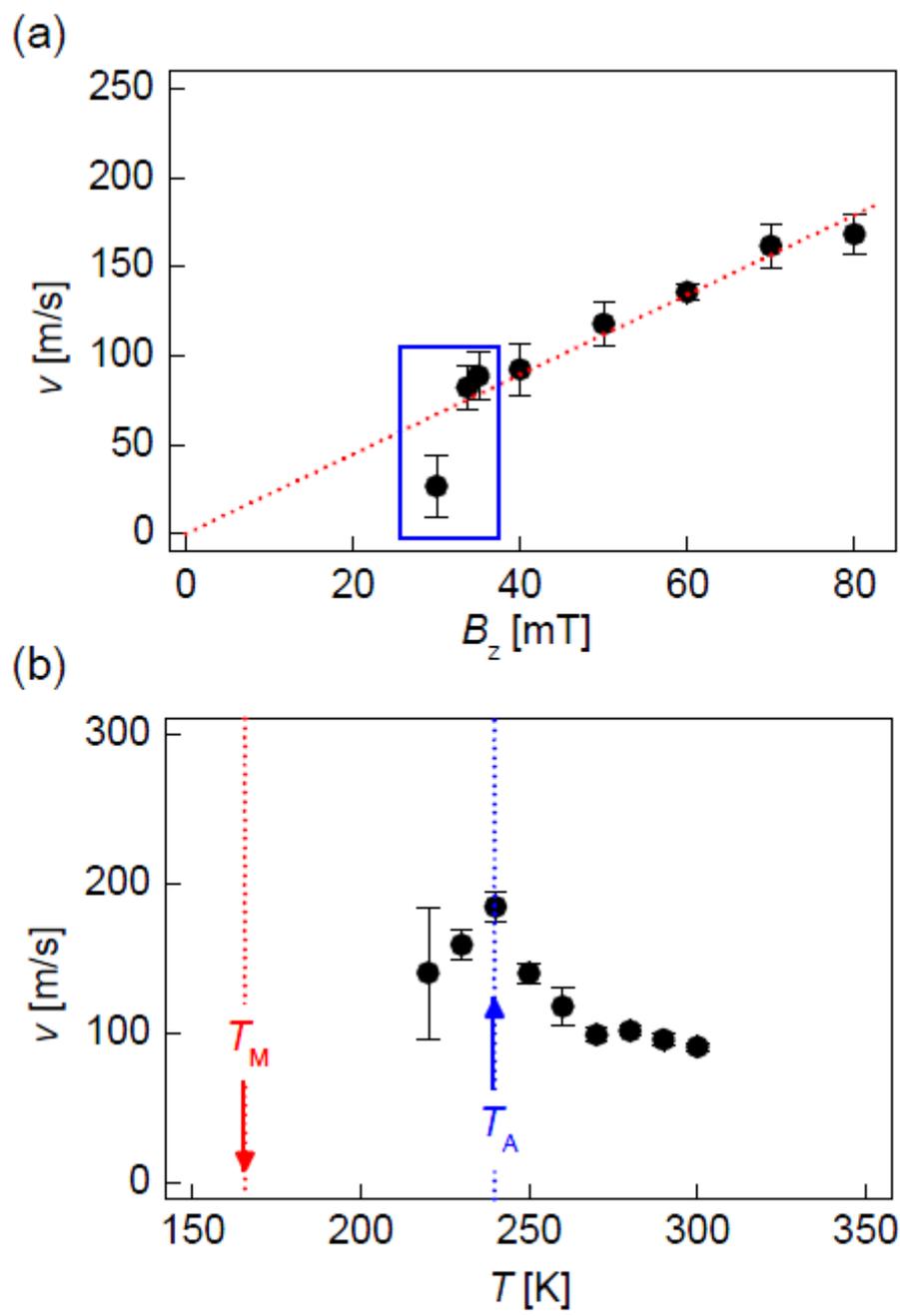

Fig. 2

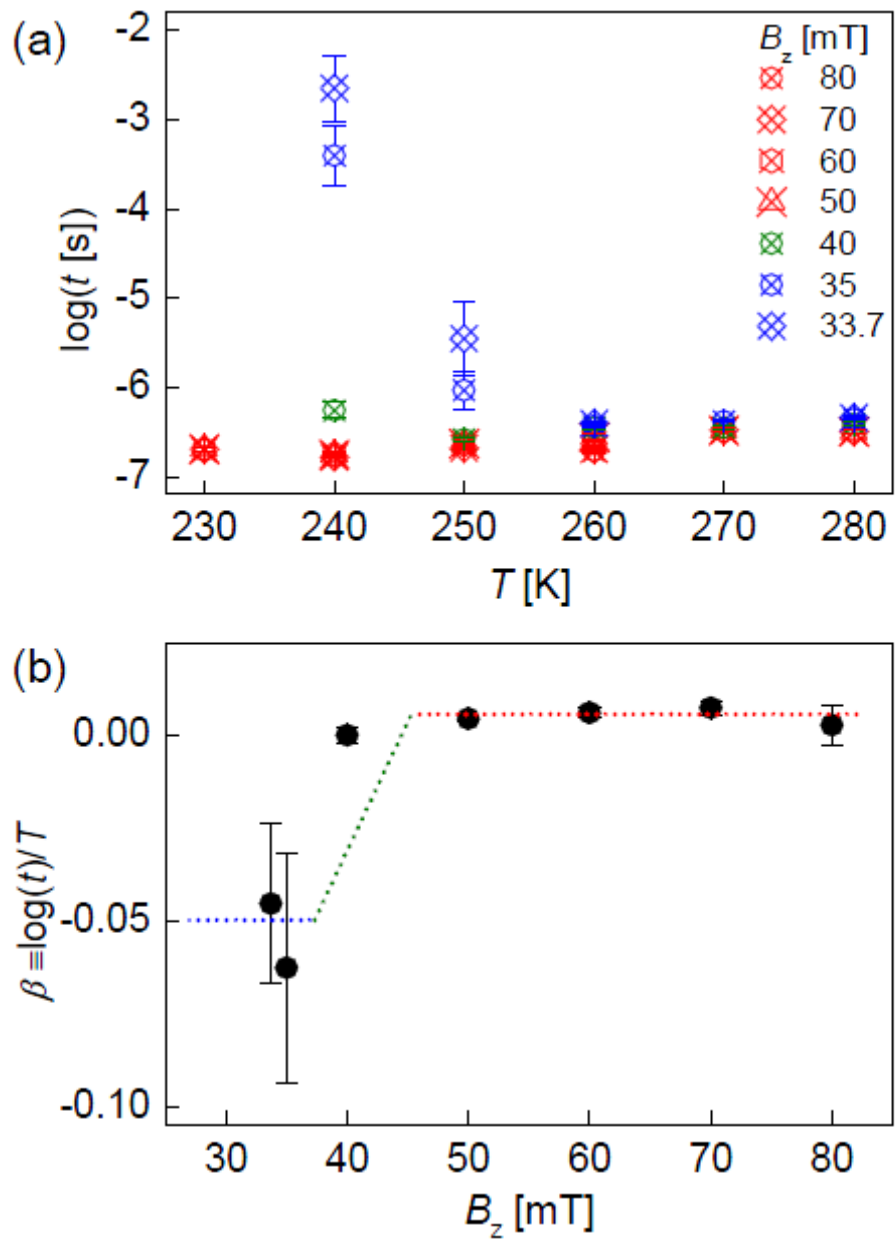

Fig. 3